\PassOptionsToPackage{unicode}{hyperref}
\PassOptionsToPackage{hyphens}{url}
\PassOptionsToPackage{dvipsnames,svgnames,x11names}{xcolor}
\documentclass[
  12pt]{article}
  
\usepackage{amsmath,amssymb,amsthm}
\usepackage{psfrag,epsf}
\usepackage{enumitem}
\usepackage{url}
\usepackage{bm}
\usepackage{hyperref}
\usepackage{threeparttable}
\usepackage{multirow}
\usepackage{tabularx}
\usepackage{xr-hyper}
\newtheorem{theorem}{Theorem}
\newtheorem{proposition}{Proposition}

\newtheorem{assumption}{Assumption}
\theoremstyle{remark}
\newtheorem{remark}{Remark}
\theoremstyle{plain}
\usepackage[ruled,vlined,linesnumbered]{algorithm2e}

\newcommand{\norm}[1]{\left\|#1\right\|}
\newcommand{\calP}{\mathcal P}

\newcommand{\dist}{\operatorname{dist}}

\newcommand{\supp}{\operatorname{supp}}

\newcommand{\spr}{\operatorname{spr}}

\usepackage{iftex}
\ifPDFTeX
  \usepackage[T1]{fontenc}
  \usepackage[utf8]{inputenc}
  \usepackage{textcomp} 
\else 
  \usepackage{unicode-math}
  \defaultfontfeatures{Scale=MatchLowercase}
  \defaultfontfeatures[\rmfamily]{Ligatures=TeX,Scale=1}
\fi
\usepackage{lmodern}
\ifPDFTeX\else  
\fi
\IfFileExists{upquote.sty}{\usepackage{upquote}}{}
\IfFileExists{microtype.sty}{
  \usepackage[]{microtype}
  \UseMicrotypeSet[protrusion]{basicmath} 
}{}
\makeatletter
\@ifundefined{KOMAClassName}{
  \IfFileExists{parskip.sty}{%
    \usepackage{parskip}
  }{
    \setlength{\parindent}{0pt}
    \setlength{\parskip}{6pt plus 2pt minus 1pt}}
}{
  \KOMAoptions{parskip=half}}
\makeatother
\usepackage{xcolor}
\makeatletter
\ifx\paragraph\undefined\else
  \let\oldparagraph\paragraph
  \renewcommand{\paragraph}{
    \@ifstar
      \xxxParagraphStar
      \xxxParagraphNoStar
  }
  \newcommand{\xxxParagraphStar}[1]{\oldparagraph*{#1}\mbox{}}
  \newcommand{\xxxParagraphNoStar}[1]{\oldparagraph{#1}\mbox{}}
\fi
\ifx\subparagraph\undefined\else
  \let\oldsubparagraph\subparagraph
  \renewcommand{\subparagraph}{
    \@ifstar
      \xxxSubParagraphStar
      \xxxSubParagraphNoStar
  }
  \newcommand{\xxxSubParagraphStar}[1]{\oldsubparagraph*{#1}\mbox{}}
  \newcommand{\xxxSubParagraphNoStar}[1]{\oldsubparagraph{#1}\mbox{}}
\fi
\makeatother

\usepackage{longtable,booktabs,array}
\usepackage{calc} 
\usepackage{etoolbox}
\makeatletter
\patchcmd\longtable{\par}{\if@noskipsec\mbox{}\fi\par}{}{}
\makeatother
\IfFileExists{footnotehyper.sty}{\usepackage{footnotehyper}}{\usepackage{footnote}}
\makesavenoteenv{longtable}
\usepackage{graphicx}
\makeatletter
\def\maxwidth{\ifdim\Gin@nat@width>\linewidth\linewidth\else\Gin@nat@width\fi}
\def\maxheight{\ifdim\Gin@nat@height>\textheight\textheight\else\Gin@nat@height\fi}
\makeatother
\setkeys{Gin}{width=\maxwidth,height=\maxheight,keepaspectratio}
\makeatletter
\def\fps@figure{htbp}
\makeatother

\makeatletter
\@ifpackageloaded{caption}{}{\usepackage{caption}}
\AtBeginDocument{%
\ifdefined\contentsname
  \renewcommand*\contentsname{Table of contents}
\else
  \newcommand\contentsname{Table of contents}
\fi
\ifdefined\listfigurename
  \renewcommand*\listfigurename{List of Figures}
\else
  \newcommand\listfigurename{List of Figures}
\fi
\ifdefined\listtablename
  \renewcommand*\listtablename{List of Tables}
\else
  \newcommand\listtablename{List of Tables}
\fi
\ifdefined\figurename
  \renewcommand*\figurename{Figure}
\else
  \newcommand\figurename{Figure}
\fi
\ifdefined\tablename
  \renewcommand*\tablename{Table}
\else
  \newcommand\tablename{Table}
\fi
}
\@ifpackageloaded{float}{}{\usepackage{float}}
\floatstyle{ruled}
\@ifundefined{c@chapter}{\newfloat{codelisting}{h}{lop}}{\newfloat{codelisting}{h}{lop}[chapter]}
\floatname{codelisting}{Listing}

\makeatother
\makeatletter
\@ifpackageloaded{caption}{}{\usepackage{caption}}
\@ifpackageloaded{subcaption}{}{\usepackage{subcaption}}
\makeatother

\ifLuaTeX
  \usepackage{selnolig}  
\fi
\usepackage[]{natbib}
\usepackage{bookmark}

\IfFileExists{xurl.sty}{\usepackage{xurl}}{} 
\hypersetup{
  pdftitle={Title},
  pdfauthor={Author 1; Author 2},
  pdfkeywords={3 to 6 keywords, that do not appear in the title},
  colorlinks=true,
  linkcolor={blue},
  filecolor={Maroon},
  citecolor={Blue},
  urlcolor={Blue},
  pdfcreator={LaTeX via pandoc}}

\newcommand{\anon}{1}

\begin{document}

\def\spacingset#1{\renewcommand{\baselinestretch}%
{#1}\small\normalsize} \spacingset{1}


\if1\anon
{
  \title{\bf A Unified Coordinate Descent Framework for Penalized Longitudinal Log-Contrast Quantile Regression}
  \author{Changwei Du, Linhui Li, Huijuan Ma\thanks{
    The authors gratefully acknowledge the Fundamental and Interdisciplinary Disciplines Breakthrough Plan of the Ministry of Education of China (JYB2025XDXM904), the National Key R\&D Program of China (2021YFA1000100, 2021YFA1000101, 2021YFA1000104), and the National Natural Science Foundation of China (12471253, 72331005).}\hspace{.2cm}\\
    School of Statistics, East China Normal University\\
    and \\
    Qi Zheng \\
    Department of Bioinformatics and Biostatistics, University of Louisville}
  \maketitle
} \fi

\if0\anon
{
  \bigskip
  \bigskip
  \bigskip
  \begin{center}
    {\LARGE\bf  A Unified Coordinate Descent Framework for Penalized Longitudinal Log-Contrast Quantile Regression}
\end{center}
  \medskip
} \fi

\bigskip
\begin{abstract}
Quantile regression characterizes heterogeneous covariate effects across the response distribution, but its extension to high-dimensional longitudinal compositional data poses substantial challenges, such as the nonsmooth check loss,  the unit-sum structure of compositional covariates. Existing work  is confined to Lasso penalization, and does not readily extend to nonconvex penalties. We propose \textit{CDaLCqr}, a unified coordinate descent algorithm for penalized longitudinal log-contrast quantile regression accommodating popular penalties, such as  adaptive Lasso, smoothly clipped absolute deviation (SCAD), and minimax concave penalty (MCP) within a single framework. The proposed method uses a Huber-type smooth approximation of the check loss and enforces the zero-sum constraint through an augmented Lagrangian formulation. \textit{CDaLCqr} combines coordinate descent with majorization–minimization, yielding explicit updates through coordinate-wise quadratic surrogates. We show that the stopping tolerance controls errors in the zero-sum constraint and stationarity conditions. Under suitable conditions and sufficiently close initialization, we also prove local convergence of the coefficient and multiplier iterates, with an error bound decreasing by a fixed factor less than one each iteration. 
Simulations highlight the estimation advantages of SCAD and MCP variants and their sparse model identification. An application to Chinese provincial industrial data identifies interpretable, quantile-specific associations between sectoral composition and regional economic output.
\end{abstract}

\noindent%
\vfill

\newpage
\spacingset{1.8} 

\section{Introduction}\label{sec-intro}

Modern longitudinal applications increasingly involve high-dimensional compositional covariates, whose components represent proportions of a whole and therefore sum to one at each observation time. In gut microbiome studies, for example, patients may be followed over multiple clinical visits, with the relative abundances of hundreds or thousands of microbial taxa jointly forming a high-dimensional compositional vector at each visit~\citep{ChenLi2016,GloballyALQR2023}. Similar data structures arise in regional economic studies, where the proportions of different industrial sectors are observed across regions and years~\citep{WangEtAl2019}. In such applications, the number of compositional covariates may substantially exceed the number of subjects. These features motivate regression methods that account for both repeated measurements and the relative nature of compositional covariates \citep{Li2015,WangEtAl2019}.

A fundamental limitation of conventional regression analyses in this domain is the reliance on conditional mean modeling, which implicitly presumes that covariate effects operate uniformly across the entire response distribution. In complex biological and socioeconomic processes, however, this assumption rarely holds and covariate effects often exhibit substantial heterogeneity across different parts of the conditional response distribution. For instance, \cite{LiEtAl2024} showed that specific microbial taxa or industrial sectors may exert minimal influence near the center of the distribution but display profound, non-negligible associations at the lower or upper extremes. Because conditional mean regression collapses these distributional nuances into a single average effect, it inherently obscures critical tail dynamics, underestimates risk factors concentrated in extreme quantiles, and can produce misleading conclusions when data exhibit heavy tails, skewness, or heteroscedasticity. Quantile regression provides an indispensable alternative by mapping how covariate effects shift across the entire conditional distribution without imposing parametric error distributions, thereby offering a complete and nuanced portrait of the underlying relationships. For longitudinal data, longitudinal quantile regression further accounts for dependence among repeated observations within subjects~\citep{Koenker2004,WangFygenson2009,Sun2016,Cho2016}. Extensions to high-dimensional longitudinal settings have employed penalization for simultaneous coefficient estimation and variable selection~\citep{GaoLiu2019}. However, these methods generally treat the covariates as unconstrained and therefore do not accommodate the unit-sum structure of compositional covariates. For compositional predictors, log-contrast regression accounts for the unit-sum structure through logarithmic transformations~\citep{AJ1982,AJBS1984}, while its symmetric representation imposes a zero-sum coefficient constraint, avoiding the choice of a reference component and providing interpretable relative effects~\citep{Lin2014,Shi2016,Lu2019}. Although sparse log-contrast methods have been developed for high-dimensional compositional covariates, they have primarily focused on mean regression or cross-sectional data. Consequently, methods that simultaneously address longitudinal quantile regression, high-dimensional variable selection, and compositional constraints remain largely undeveloped.

Simultaneously incorporating these features introduces both statistical and computational hurdles. 
Within-subject temporal dependence complicates variance structures, while the unit-sum constraint implies that components cannot vary independently. Directly incorporating such covariates into conventional regression models can yield misleading, baseline-dependent interpretations. 
 High dimensionality, where the number of compositional covariates may far exceed the sample size, necessitates variable selection and regularization, which in turn increases the computational burden. In addition, the check loss used in quantile regression is non-differentiable at the origin, making direct optimization difficult. Together, these issues imply that standard algorithms developed for unconstrained high-dimensional quantile regression are not directly applicable.

To date, the primary framework directly tackling this unified setting is \citet{GloballyALQR2023}, which developed a globally adaptive longitudinal quantile regression method for high-dimensional compositional covariates. Their procedure combines a longitudinal quantile loss with an adaptive Lasso penalty subject to the zero-sum coefficient constraint and establishes uniform estimation and model selection properties over a range of quantile levels.  For computation, the penalized problem is reformulated as a constrained quantile regression problem and solved via the linear programming routine \texttt{rq.fit.fnc()} in R. However, because this construction relies fundamentally on representing the $\ell_1$ norm via augmented pseudo-observations within the check loss,  it is specific to the adaptive Lasso and cannot be extended directly to the smoothly clipped absolute deviation (SCAD) penalty or the minimax concave penalty (MCP). Although adaptive weighting alleviates some of the shrinkage associated with the ordinary Lasso, it may still excessively penalize large coefficients in finite samples~\citep{Zou01122006,PotscherSchneider2009}. Nonconvex penalties such as SCAD and MCP reduce the penalization imposed on sufficiently large signals and may consequently improve coefficient estimation and variable selection accuracy.

Penalized quantile regression under SCAD or MCP is typically tackled via coordinate descent, semismooth Newton methods, or alternating direction schemes~\citep{PengWang2015,Mkhadri2017,YiHuang2017,ADMM2018}. Among these, coordinate descent is ideal for high-dimensional regimes because it replaces a large optimization problem with a sequence of inexpensive one-dimensional updates and can naturally exploit coefficient sparsity. In the present setting, however, coordinate descent faces two severe obstacles: the non-differentiable check loss precludes gradient-based coordinate steps, while the zero-sum constraint fundamentally couples all compositional coefficients, preventing standard coordinate-wise decoupling. These complications, paired with the potential non-convexity of SCAD and MCP, create an optimization problem that existing coordinate algorithms cannot directly solve.
A unified optimization scheme is therefore required to decouple the constraint and manage diverse penalties within a single framework.

To address these challenges, we propose \textit{CDaLCqr}, a unified coordinate descent algorithm for penalized longitudinal log-contrast quantile regression. In this framework, adaptive Lasso, SCAD, and MCP are handled through a common augmented Lagrangian formulation and coordinate descent scheme. We replace the non-differentiable check loss with a Huber-type smooth approximation, which is continuously differentiable and admits gradient-based optimization. The repeated-measurement structure is accommodated by formulating the objective as a longitudinal quantile loss, which sums the smoothed check losses over all observed follow-up times within each subject. We then incorporate the zero-sum constraint through an augmented Lagrangian formulation, converting the constrained problem into a sequence of unconstrained subproblems. The core of the algorithm is a coordinate-wise majorization--minimization (MM) scheme in which the smooth loss is replaced at each coordinate step by a quadratic surrogate that majorizes the smooth loss and yields explicit thresholding updates for the penalties. The key to the proposed algorithm is a quadratic augmentation term in the augmented Lagrangian, which provides additional curvature that restores convexity of each coordinate subproblem, even under nonconvex penalties. 

The proposed framework makes several contributions. First, it establishes a unified coordinate descent optimization scheme that accommodates adaptive Lasso, SCAD, and MCP through explicit, closed-form thresholding updates. Second, it provides rigorous finite-termination error bounds on the constraint violation and stationarity conditions, and establishes local linear convergence of the coefficient and multiplier sequences under regularity and spectral stability conditions governing the contractive properties of the algorithmic mapping. Third, simulation studies demonstrate favorable estimation and variable selection performance, with the SCAD and MCP variants yielding lower mean estimation errors than the adaptive Lasso approach in most settings. We further evaluate the empirical iteration matrix $\widehat T$ at the numerical termination point of each fit and find that the local spectral stability condition is satisfied in 99.67\% of the simulation fits. Finally, an empirical application to Chinese provincial industrial structure data demonstrates the practical utility of the method, uncovering varying sectoral contributions to regional economic output across different conditional quantiles.

The remainder of this paper is organized as follows. Section~\ref{methodology} introduces the longitudinal log-contrast quantile regression model and develops the proposed algorithm. It establishes convexity of the coordinate objectives, bounds the errors at termination, and gives conditions for local convergence, with proofs in Section~\ref{app:convergence} of the supplementary material. Section~\ref{sec:simu} presents simulation studies evaluating the finite-sample performance of the method. Section~\ref{sec:empi} applies the method to Chinese provincial industrial structure data. Section~\ref{sec:con&dis} concludes with a brief discussion.

\section{Methodology}
\label{methodology}

This section develops \textit{CDaLCqr} for longitudinal quantile regression with high-dimensional compositional covariates. We first formulate longitudinal log-contrast quantile regression with a zero-sum constraint and introduce adaptive Lasso, SCAD, and MCP regularization. We then smooth the check loss and incorporate the zero-sum constraint through an augmented Lagrangian formulation. Next, we construct a coordinate-wise majorization--minimization algorithm and derive explicit updates for all model parameters. Finally, we establish the strict convexity of the coordinate objectives, bound the errors of the zero-sum constraint and the stationarity conditions at termination, and give conditions for local convergence to a stationary point and its associated multiplier.

\subsection{Longitudinal log-contrast quantile regression model}

Consider a longitudinal study with $n$ subjects. Let $Y_{i}(t)$, $X_{i}(t)$, and $W_{i}(t)$ denote the longitudinal outcome, the $r$-dimensional covariate process vector (with 1 as the first component for the intercept), and the $p$-dimensional vector of compositional covariates for the $i$th subject at time point $t$, respectively, $i = 1,\ldots,n$. At each time point $t$, the compositional covariate $W_{i}(t)$ is subject to the unit-sum constraint, hence it lies in the $(p - 1)$-dimensional positive simplex $$S^{p - 1} = \{ \left( w_{1},\ldots,w_{p} \right):w_{j} > 0,j = 1,\ldots,p;\sum_{j = 1}^{p}{w_{j} = 1} \}.$$ The variables $Y_{i}(t)$, $X_{i}(t)$, $W_{i}(t)$ are observed at a sequence of follow-up times $\{ t_{iv},v = 1,\ldots,K_{i} \}$, where $K_{i}$ is the total number of follow-ups for the $i$th subject, and the counting process for the follow-up is denoted as $ N_i(t)=\sum_{v=1}^{K_i}I(t_{iv}\le t)$.

Given $X_i(t)$ and $W_i(t)$, the $\tau$-conditional quantile function of $Y_i(t)$ is defined as:
$$Q_{Y_i(t)}\{ \tau | X_i(t),W_i(t) \} \doteq \inf\{ y:\Pr{\{ Y_i(t) \leq y|X_i(t),W_i(t) \} \geq \tau}\}.$$
Directly incorporating $W_i(t)$ into a regression model is problematic because its components cannot vary freely under the unit-sum constraint, rendering individual regression coefficients uninterpretable. Reference-based log-ratio models provide equivalent parameterizations of the zero-sum linear predictor. However, applying coordinatewise penalties directly to reference-specific coefficients can make the penalized estimator depend on the selected reference. Following \citet{Lin2014}, we therefore adopt the symmetric representation
\[
Z_i(t)=\left(\log W_{i1}(t),\ldots,\log W_{ip}(t)\right)^{T},
\]
and consider
\[
Q_{Y_i(t)}\left(\tau\mid X_i(t),Z_i(t)\right)
=
X_i(t)^T\alpha_0(\tau)+Z_i(t)^T\beta_0(\tau),
\qquad
\sum_{j=1}^{p}\beta_{0j}(\tau)=0.
\]
The zero-sum constraint $\sum_{j=1}^p \beta_{0j}(\tau) = 0$ guarantees invariance to the choice of reference component and treats all compositional elements symmetrically.

For a fixed quantile level $\tau$, we define the longitudinal quantile loss under working independence as:
\begin{equation}
Q(\alpha,\beta;\tau)
=
\frac{1}{n}
\sum_{i=1}^{n}
\int_{0}^{\infty}
\rho_\tau
\left\{
Y_i(t)-X_i(t)^T\alpha(\tau)-Z_i(t)^T\beta(\tau)
\right\}
dN_i(t),
\label{eq:quantile-loss}
\end{equation}
where $\rho_\tau(u)=u\{\tau-I(u\leq0)\}$ is the standard check loss. This
working-independence formulation aggregates the quantile losses over all observation times without explicitly modeling the within-subject
dependence structure.

To enforce sparsity and perform variable selection under the zero-sum constraint, we consider the penalized estimator:
\begin{equation}
\min_{\alpha,\beta}
\left\{
Q(\alpha,\beta;\tau)
+
\sum_{j=1}^{p}P_\lambda\bigl(|\beta_j(\tau)|\bigr)
\right\},
\qquad \text{subject to} \qquad
\sum_{j=1}^{p}\beta_j(\tau)=0,
\label{eq:penalized-problem}
\end{equation}
where $\lambda>0$ is the tuning parameter and $P_\lambda(\cdot)$ is chosen
from the adaptive Lasso, MCP, and SCAD penalties.

For the adaptive Lasso,
\[
P_\lambda\bigl(|\beta_j(\tau)|\bigr)
=
\lambda\omega_j|\beta_j(\tau)|,
\]
where $\omega_j>0$ 
are data-driven adaptive weights derived from a preliminary estimator. When this penalty is chosen,  \eqref{eq:penalized-problem} coincides with the formulation of \citet{GloballyALQR2023} at a fixed quantile
level.

For MCP with $a>1$,
\[
P_\lambda(u)
=
\begin{cases}
\lambda u-\dfrac{u^2}{2a},
& 0\leq u\leq a\lambda,\\[6pt]
\dfrac{a\lambda^2}{2},
& u>a\lambda.
\end{cases}
\]

For SCAD with $a>2$,
\[
P_\lambda(u)
=
\begin{cases}
\lambda u,
& 0\leq u\leq\lambda,\\[4pt]
\dfrac{2a\lambda u-u^2-\lambda^2}{2(a-1)},
& \lambda<u\leq a\lambda,\\[8pt]
\dfrac{(a+1)\lambda^2}{2},
& u>a\lambda.
\end{cases}
\]
Because the check loss $\rho_\tau(u)$ is non-differentiable at the origin, gradient-based optimization cannot be applied directly. We replace $\rho_\tau$ with the Huber-type smooth approximation $\rho_{\tau,k}$:
\begin{equation}
Q_k(\alpha,\beta;\tau)
=
\frac{1}{n}
\sum_{i=1}^{n}
\int_{0}^{\infty}
\rho_{\tau,k}
\left\{
Y_i(t)-X_i(t)^T\alpha(\tau)-Z_i(t)^T\beta(\tau)
\right\}
dN_i(t),
\label{eq:smooth-loss}
\end{equation}
where
\[
\rho_{\tau,k}(u)
=
\begin{cases}
(\tau-1)u-\dfrac{k(1-\tau)^2}{2},
& u<-(1-\tau)k,\\[8pt]
\dfrac{u^2}{2k},
& -(1-\tau)k\leq u\leq \tau k,\\[8pt]
\tau u-\dfrac{k\tau^2}{2},
& u>\tau k.
\end{cases}
\]

\begin{remark}
More generally, any convex continuously differentiable approximation with a Lipschitz continuous derivative may be utilized.
For example, adapting the asymmetric quadratic smoothing of
\citet{jennings1996optimal} to the check loss gives
\begin{equation*}
\rho_{\tau,\delta}(u)
=
\rho_\tau(u)
+
\frac{
\bigl(\delta-\tau u_+\bigr)_+^2
+
\bigl\{\delta-(1-\tau)(-u)_+\bigr\}_+^2
}{2\delta}
-
\frac{\delta}{2},
\label{eq:jennings-smoothing}
\end{equation*}
where $u_+=\max(u,0)$ and $\delta>0$ is a smoothing parameter. The corresponding derivative
is Lipschitz continuous, with Lipschitz constant
$\max\{\tau^2,(1-\tau)^2\}/\delta$. Throughout this paper, derivations are presented using the Huber-type approximation $\rho_{\tau,k}$.
\end{remark}

Replacing $Q$ with $Q_k$ removes the non-differentiability of the loss, but the zero-sum constraint on the compositional coefficients remains to be enforced. To handle the zero-sum constraint without destroying coordinate separability, we formulate the augmented Lagrangian objective: \begin{equation} \begin{aligned} G(\alpha,\beta,\gamma;\tau) =Q_k(\alpha,\beta;\tau) + \sum_{j=1}^{p}P_\lambda\bigl(|\beta_j(\tau)|\bigr)+ \gamma\sum_{j=1}^{p}\beta_j(\tau) + \frac{\mu}{2} \left\{\sum_{j=1}^{p}\beta_j(\tau)\right\}^{2}, \end{aligned} \label{eq:augmented-objective} \end{equation} where $\gamma$ is the Lagrange multiplier and $\mu>0$ is the augmentation parameter. The linear term $\gamma\sum_{j=1}^{p}\beta_j(\tau)$ is the usual Lagrange multiplier term associated with the zero-sum constraint, while the quadratic term $\frac{\mu}{2} \{\sum_{j=1}^{p}\beta_j(\tau)\}^{2}$ penalizes the violation of that constraint and improves the numerical stability of the constrained optimization problem. When $\sum_{j=1}^{p}\beta_j(\tau)=0$, both terms are equal to zero, and the augmented objective reduces to the original smoothed penalized objective.

\subsection{Coordinate descent and majorization for the compositional coefficients}

Although the augmented objective in \eqref{eq:augmented-objective} provides a convenient unconstrained formulation for fixed $\gamma$, its minimization over all components of $\alpha(\tau)$ and $\beta(\tau)$ remains computationally demanding in high-dimensional settings. We therefore adopt a coordinate descent algorithm (CDA), updating one coefficient at a time while holding the remaining coefficients fixed. Within each coordinate iteration, the newly updated value of a coordinate is used immediately in the subsequent updates.

Exact minimization of each coordinate-wise objective is still inconvenient because the loss does not generally lead to a simple closed-form solution. To obtain tractable updates, we embed each coordinate step in a majorization--minimization (MM) scheme. At the current iterate, the smooth loss is replaced by a quadratic surrogate that majorizes the original loss and agrees with it at the current value. Minimizing this surrogate therefore gives a non-increasing coordinate update of the augmented objective. We first derive the update for $\beta_j(\tau)$ because the compositional coefficients are subject to penalization and variable selection. The unpenalized update for $\alpha_l(\tau)$ follows from the same argument after removing the penalty and constraint terms.

Let $\widetilde\alpha(\tau)$ and $\widetilde\beta(\tau)$ denote the working coefficient vectors immediately before a coordinate update, and let $\widetilde\gamma$ denote the current value of the Lagrange multiplier. We hold $\widetilde\gamma$ fixed while updating the coefficients and update the multiplier once all coordinates have been visited. To derive the update for $\beta_j(\tau)$, define the working
residual
\begin{equation*}
\widetilde r_i(t)
=
Y_i(t)-X_i(t)^T\widetilde\alpha(\tau)
-Z_i(t)^T\widetilde\beta(\tau).
\label{eq:working_residual}
\end{equation*}
If the current value $\widetilde\beta_j(\tau)$ is replaced by a candidate value
$b$, the corresponding residual becomes
\[
\widetilde r_i(t)
+
Z_{ij}(t)\{\widetilde\beta_j(\tau)-b\}.
\]
The penalties on the other compositional coefficients and their contribution to the linear multiplier term sum to $\sum_{l\ne j}P_\lambda\bigl(|\widetilde\beta_l(\tau)|\bigr) +\widetilde\gamma\sum_{l\ne j}\widetilde\beta_l(\tau).$ This quantity is constant with respect to $b$, so omitting it preserves the minimizer. The coordinate-wise objective is
\begin{multline}
G_{\beta_j}
\left\{b\mid\widetilde\alpha(\tau),\widetilde\beta(\tau)\right\}
=
\frac{1}{n}\sum_{i=1}^{n}
\int_{0}^{\infty}
\rho_{\tau,k}
\left[
\widetilde r_i(t)
+
Z_{ij}(t)\{\widetilde\beta_j(\tau)-b\}
\right]
\,dN_i(t) \\
+
P_\lambda(|b|)
+
\widetilde\gamma b
+
\frac{\mu}{2}
\left\{
 b+\sum_{l\ne j}\widetilde\beta_l(\tau)
\right\}^{2}.
\label{eq:beta_coordinate_objective}
\end{multline}

The working vector $\widetilde\beta(\tau)$ is updated sequentially, with each coordinate update using the latest available values.

Since $\rho_{\tau,k}'(\cdot)$ is Lipschitz continuous with constant $1/k$ \citep{Mkhadri2017}, the smoothed loss satisfies the quadratic majorization bound:
   $$\rho_{\tau,k}(u - v) \le \rho_{\tau,k}(u) - \rho_{\tau,k}'(u)v + \frac{1}{2k}v^2.$$
Setting $u = \tilde{r}_i(t)$ and $v = Z_{ij}(t)\{b - \tilde{\beta}_j(\tau)\}$, we construct the quadratic surrogate:
\begin{multline}
F_{\beta_j}
\left\{b\mid\widetilde\alpha(\tau),\widetilde\beta(\tau)\right\}
=
\frac{1}{n}\sum_{i=1}^{n}
\int_{0}^{\infty}
\left[
\rho_{\tau,k}\{\widetilde r_i(t)\}
+
\rho_{\tau,k}'\{\widetilde r_i(t)\}Z_{ij}(t)
\{\widetilde\beta_j(\tau)-b\}
\right. \\
\left.
+
\frac{1}{k}Z_{ij}^{2}(t)
\{\widetilde\beta_j(\tau)-b\}^{2}
\right]
\,dN_i(t)
+
P_\lambda(|b|)
+
\widetilde\gamma b
+
\frac{\mu}{2}
\left\{
 b+\sum_{l\ne j}\widetilde\beta_l(\tau)
\right\}^{2}.
\label{eq:beta_surrogate}
\end{multline}
The surrogate agrees with the coordinate-wise objective at the current value,
\[
F_{\beta_j}
\left\{\widetilde\beta_j(\tau)\mid
\widetilde\alpha(\tau),\widetilde\beta(\tau)\right\}
=
G_{\beta_j}
\left\{\widetilde\beta_j(\tau)\mid
\widetilde\alpha(\tau),\widetilde\beta(\tau)\right\},
\]
and $F_{\beta_j}(b) \ge G_{\beta_j}(b)$ for all $b$. Hence, updating $b$ by minimizing $F_{\beta_j}$ guarantees the descent property for $G_{\beta_j}$. The MM property and the coordinate optimality condition are established in Lemma~\ref{conv:lem:update} in Section~\ref{app:convergence} of the supplementary material.

At this point, the second role of the quadratic augmentation becomes clear. As a function of the candidate coordinate $b$, the term
$\frac{\mu}{2}
\{b+\sum_{l\ne j}\widetilde\beta_l(\tau)\}^{2}
$
contributes positive curvature $\mu$ to both the coordinate-wise objective and its surrogate. This additional curvature offsets the concave components of the SCAD and MCP penalties and can therefore restore convexity of the resulting one-dimensional minimization problem. The following proposition formalizes this property.

\begin{proposition}
\label{thm:coordinate_convexity}
Consider the coordinate-wise objective $G_{\beta_j}\{\cdot\mid\widetilde\alpha(\tau),\widetilde\beta(\tau)\}$ in \eqref{eq:beta_coordinate_objective} and its MM surrogate $F_{\beta_j}\{\cdot\mid\widetilde\alpha(\tau),\widetilde\beta(\tau)\}$ in \eqref{eq:beta_surrogate}. For the adaptive Lasso, MCP $(a>1)$, and SCAD $(a>2)$ penalties considered in this paper, both $G_{\beta_j}\{b\mid\widetilde\alpha(\tau),\widetilde\beta(\tau)\}$ and $F_{\beta_j}\{b\mid\widetilde\alpha(\tau),\widetilde\beta(\tau)\}$ are convex $b$ whenever $\mu\geq1$.
\end{proposition}

Accordingly, we take $\mu\geq1$ in the algorithm and convergence analysis below. The proof of Proposition~\ref{thm:coordinate_convexity} is provided in Section~\ref{app:convergence} of the supplementary material.

\subsection{Coordinate descent updates}

Let $\alpha^{(m)}(\tau)$, $\beta^{(m)}(\tau)$, and $\gamma^{(m)}$ denote the two coefficient vectors and the Lagrange multiplier at the start of $m$th iteration. We set $\widetilde\alpha(\tau)=\alpha^{(m)}(\tau)$, $\widetilde\beta(\tau)=\beta^{(m)}(\tau)$, and $\widetilde\gamma=\gamma^{(m)}$.

We now derive the explicit coordinate-wise updates obtained by minimizing the surrogate $F_{\beta_j}\{b \mid \tilde{\alpha}(\tau), \tilde{\beta}(\tau)\}$ in \eqref{eq:beta_surrogate}. Expanding the quadratic terms in $b$ and grouping terms, the surrogate objective can be written in canonical one-dimensional thresholding form up to constants independent of $b$:
   $$F_{\beta_{j}}\{b \mid \tilde{\alpha}(\tau),\tilde{\beta}(\tau)\} = \frac{1}{2}\zeta_{2j} b^2 - \zeta_{1j}^{(m)} b + P_{\lambda}(\vert{}b\vert{}) + \text{const},$$
   where the coordinate-specific coefficients are given by:
\begin{equation}
\begin{aligned}
\zeta_{1j}^{(m)}
=&
\left\{
\frac{2}{kn}\sum_{i=1}^n\int_0^\infty Z_{ij}^2(t)\,dN_i(t)
\right\}\widetilde\beta_j(\tau)
+
\frac{1}{n}\sum_{i=1}^n\int_0^\infty
\rho'_{\tau,k}\{\widetilde r_i(t)\}Z_{ij}(t)\,dN_i(t) \\
&-
\mu\sum_{l\ne j}\widetilde\beta_l(\tau)
-
\gamma^{(m)},
\end{aligned}
\label{eq:zeta1}
\end{equation}
and
\begin{equation}
\zeta_{2j}
=
\frac{2}{kn}\sum_{i=1}^n\int_0^\infty Z_{ij}^2(t)\,dN_i(t)
+
\mu.
\label{eq:zeta2}
\end{equation}
With this notation, the minimizer of \eqref{eq:beta_surrogate} has an explicit thresholding form using the standard soft thresholding rule \citep{JSSv033i01}.

For the adaptive Lasso,
\begin{equation}
\beta_j^{(m+1)}(\tau)
=
\frac{S(\zeta_{1j}^{(m)},\lambda\omega_j)}
{\zeta_{2j}},
\qquad
\label{eq:adaptive_lasso_update}
\end{equation}
where $S(z,u)=\operatorname{sgn}(z)(|z|-u)_+$ denotes the soft thresholding operator.

For MCP, the coordinate update is
\begin{equation}
\beta_j^{(m+1)}(\tau)
=
\begin{cases}
\dfrac{S(\zeta_{1j}^{(m)},\lambda)}
{\zeta_{2j}-\frac{1}{a}},
& |\zeta_{1j}^{(m)}|\le a\lambda \zeta_{2j},\\[10pt]
\dfrac{\zeta_{1j}^{(m)}}{\zeta_{2j}},
& |\zeta_{1j}^{(m)}|>a\lambda \zeta_{2j}.
\end{cases}
\label{eq:mcp_update}
\end{equation}

For SCAD, the coordinate update is
\begin{equation}
\beta_j^{(m+1)}(\tau)
=
\begin{cases}
\dfrac{S(\zeta_{1j}^{(m)},\lambda)}
{\zeta_{2j}},
& |\zeta_{1j}^{(m)}|\le \lambda(1+\zeta_{2j}),\\[10pt]
\dfrac{
S\left(\zeta_{1j}^{(m)},\dfrac{a\lambda}{a-1}\right)}
{\zeta_{2j}-\frac{1}{a-1}},
& \lambda(1+\zeta_{2j})<|\zeta_{1j}^{(m)}|
\le a\lambda \zeta_{2j},\\[12pt]
\dfrac{\zeta_{1j}^{(m)}}{\zeta_{2j}},
& |\zeta_{1j}^{(m)}|>a\lambda \zeta_{2j}.
\end{cases}
\label{eq:scad_update}
\end{equation}

The unpenalized coefficients $\alpha_l(\tau) (l=1,\dots,r)$ are updated through the same quadratic
majorization principle, without the penalty and zero-sum constraint terms. Applying coordinate minimization with step size derived from the surrogate bound gives:
\begin{equation}
\alpha_l^{(m+1)}(\tau)
=
\alpha_l^{(m)}(\tau)
+
\frac{
\sum_{i=1}^n\int_0^\infty
\rho'_{\tau,k}\{\widetilde r_i(t)\}X_{il}(t)\,dN_i(t)
}{
2k^{-1}\sum_{i=1}^n\int_0^\infty X_{il}^2(t)\,dN_i(t)
}.
\label{eq:alpha_update}
\end{equation}
After cyclic updates of all unpenalized coordinates $\alpha_l$ followed by all compositional coordinates $\beta_j$, the full parameter vectors are updated as $\alpha^{(m+1)}(\tau) = \tilde{\alpha}(\tau)$ and $\beta^{(m+1)}(\tau) = \tilde{\beta}(\tau)$. The Lagrange multiplier is then updated via the first-order augmented Lagrangian step based on the current constraint infeasibility:
\begin{equation}
\gamma^{(m+1)}
=
\gamma^{(m)}
+
\mu\sum_{j=1}^p\beta_j^{(m+1)}(\tau).
\label{eq:gamma_update}
\end{equation}

\subsection{The \textit{CDaLCqr} algorithm}

The complete CDaLCqr procedure is detailed in Algorithm 1. At each outer iteration $m$, the algorithm performs a cyclic sweep over all $r$ unpenalized coefficients $\alpha_l(\tau)$ and all $p$ compositional coefficients $\beta_j(\tau)$. To minimize computational cost in high dimensions, the working residuals $\tilde{r}_i(t)$ are updated sequentially after each univariate coordinate update.   Once a complete cycle of coordinate updates is concluded, the Lagrange multiplier $\gamma^{(m)}$ is updated using the zero-sum constraint residual via \eqref{eq:gamma_update}. The outer loop ends when the maximum coordinate change across both $\alpha$ and $\beta$, together with the constraint infeasibility, drops below the tolerance $\varepsilon$, or when the iteration cap $M_{\max}$ is reached. The algorithm returns a flag indicating whether the tolerance was satisfied, the final stopping quantity, and the terminal multiplier.
\begingroup
\spacingset{1.2}
\begin{algorithm}[H]
\caption{$CDaLCqr$}
\label{alg:cdalcqr}

\KwIn{Data $\{Y_i(t),X_i(t),W_i(t)\}$, quantile level $\tau$, penalty type $\mathcal P$, tuning parameters $(\lambda,k,\mu)$ with $\mu\geq1$, tolerance $\varepsilon$, and maximum iteration number $M_{\max}$.}

\KwOut{$\widehat\alpha(\tau)$, $\widehat\beta(\tau)$, selected set $\widehat S_\tau$, terminal multiplier $\widehat\gamma$, tolerance-satisfied flag $c_{\mathrm{tol}}$, final stopping quantity $\widehat\Delta$.}

Transform compositional covariates by $Z_{ij}(t)=\log W_{ij}(t)$\;

Initialize $\alpha^{(0)}$, $\beta^{(0)}$, and $\gamma^{(0)}=0$\;

\If{$\mathcal P$ is adaptive Lasso}{
    Compute adaptive weights $\omega_j$, $j=1,\ldots,p$\;
}

\For{$m=0,1,\ldots,M_{\max}-1$}{

    Set
    $\widetilde\alpha\leftarrow\alpha^{(m)}$ and
    $\widetilde\beta\leftarrow\beta^{(m)}$\;

    Compute the residuals at the beginning of the outer iteration:
    \[
    r_i^{(m)}(t)
    =
    Y_i(t)
    -
    X_i(t)^T\alpha^{(m)}(\tau)
    -
    Z_i(t)^T\beta^{(m)}(\tau).
    \]

    Set $\widetilde r_i(t)\leftarrow r_i^{(m)}(t)$\;
    \For{$\ell=1,\ldots,r$}{
    Update $\widetilde\alpha_\ell$ according to
    \eqref{eq:alpha_update}, with the loss derivative
    evaluated at the current $\widetilde r_i(t)$\;

    $\widetilde r_i(t)\leftarrow
    \widetilde r_i(t)
    -X_{i\ell}(t)
    \{\widetilde\alpha_\ell-\alpha_\ell^{(m)}\}$\;
}

    \For{$j=1,\ldots,p$}{
        Update $\widetilde\beta_j$ using the thresholding rule
        corresponding to $\mathcal P$:
        \eqref{eq:adaptive_lasso_update},
        \eqref{eq:mcp_update}, or
        \eqref{eq:scad_update}\;  
        
        Update working residual $\widetilde r_i(t)$\;
    }

    Set
    $\alpha^{(m+1)}\leftarrow\widetilde\alpha$ and
    $\beta^{(m+1)}\leftarrow\widetilde\beta$\;

    Update $\gamma^{(m+1)}$ by \eqref{eq:gamma_update}\;

    Compute
    \[
    \Delta^{(m)}=\max \left\{ \|\alpha^{(m+1)} - \alpha^{(m)}\|_\infty, \|\beta^{(m+1)} - \beta^{(m)}\|_\infty, \left| \sum_j \beta_j^{(m+1)} \right| \right\}.
    \]
    
    \If{$\Delta^{(m)}<\varepsilon$}{
        \textbf{break}\;
    }
}

Set $\widehat\alpha(\tau)=\alpha^{(m+1)}$ 
$\widehat\beta(\tau)=\beta^{(m+1)}$, $\widehat\gamma=\gamma^{(m+1)}$, $\widehat\Delta=\Delta^{(m)}$, and
$c_{\mathrm{tol}}=\mathbf1\{\widehat\Delta<\varepsilon\}$\;

Define
$\widehat S_\tau=\{j:|\widehat\beta_j(\tau)|>0\}$\;

\Return{$\widehat\alpha(\tau)$, $\widehat\beta(\tau)$, $\widehat S_\tau$, $\widehat\gamma$, $c_{\mathrm{tol}}$, \text{and} $\widehat\Delta$}\;

\end{algorithm}
\endgroup
\subsection{Accuracy at termination and local convergence}
\label{sec:convergence}
We next examine the theoretical properties of the stopping rule and the local convergence of Algorithm 1. We first show that upon satisfaction of the stopping criterion, the terminal iterate approximately satisfies both the zero-sum constraint and the first-order stationarity conditions of the penalized problem. We then establish conditions under which the joint sequence of coefficient and multiplier iterates converges locally to a Karush–Kuhn–Tucker (KKT) stationary pair. All proofs are provided in Section S1 of the Supplementary Material.

Throughout this subsection, we fix the quantile level $\tau\in(0,1)$ and abbreviate $\alpha$ and $\beta$ as $\alpha(\tau)$ and $\beta(\tau)$, respectively. We concatenate all model parameters into $\theta = (\alpha^T, \beta^T)^{T} \in \mathbb{R}^d$, where $d = r + p$. Let $e = (0_r^T, 1_p^T)^T \in \mathbb{R}^d$, so that the zero-sum constraint on the compositional coefficients is compactly expressed as $e^{T} \theta = 0$. Stack the $N = \sum_{i=1}^n K_i$ total observations into the response vector $y \in \mathbb{R}^N$ and design matrix $V = [X \;\; Z] \in \mathbb{R}^{N \times d}$. We denote the $v$-th row of $V$ by $V_{v:}$ and the $l$-th column by $V_{:l}$. The smoothed longitudinal quantile objective and the separable penalty function are written as:
\begin{equation}
 Q_k(\theta)=\frac{1}{n  }\sum_{v=1}^N
 \rho_{\tau,k}(y_v-V_{v:}\theta),\qquad
 \calP(\theta)=\sum_{j=1}^pP_{\lambda,j}(|\beta_j|).
 \label{conv:eq:loss}
\end{equation}

All algorithmic parameters are held fixed throughout the analysis: $k > 0$, $\lambda > 0$, and $\mu \ge 1$, with shape parameters $a > 1$ for MCP and $a > 2$ for SCAD. We assume that all covariate and response entries are finite, and every column of $V$ is non-zero. Under Proposition~\ref{thm:coordinate_convexity}, the quadratic majorization surrogate minimized at each coordinate step is strictly convex and possesses a unique minimizer. 

A first-order stationary point $\theta^*$ of the constrained problem satisfies the zero-sum constraint and admits an associated Lagrange multiplier $\gamma^* \in \mathbb{R}$ satisfying the subdifferential inclusion:
\begin{equation}
 e^{T}\theta^*=0,\qquad
 0\in\nabla Q_k(\theta^*)+\partial\calP(\theta^*)+\gamma^*e.
 \label{conv:eq:kkt}
\end{equation}
where $\partial \mathcal{P}(\theta^*)$ denotes the limiting Fréchet subdifferential of $\mathcal{P}$ at $\theta^*$. For any set $\mathcal{A} \subset \mathbb{R}^d$, let $\operatorname{dist}(0, \mathcal{A}) = \inf_{a \in \mathcal{A}} \Vert{}a\Vert{}_2$ quantify the stationarity residual.   

The following theorem shows how closely the terminal iterate of Algorithm 1  satisfy these two conditions in \eqref{conv:eq:kkt} when the stopping tolerance is satisfied.

\begin{theorem}[Accuracy at termination]
\label{conv:thm:stopping}
Under the conditions stated above, suppose that an iteration of Algorithm~\ref{alg:cdalcqr} satisfies
\begin{equation}
 \Delta^{(m)}\le\varepsilon,
 \label{conv:eq:stop}
\end{equation}
where $\Delta^{(m)}$ is the stopping quantity defined in the algorithm. Then the updated coefficients and multiplier satisfy
\begin{align}
 |e^{T}\theta^{(m+1)}|
 \le\varepsilon,\quad
 \dist\left(
 0,\nabla Q_k(\theta^{(m+1)})
 +\partial\calP(\theta^{(m+1)})
 +\gamma^{(m+1)}e
 \right)
 &\le C_{\mathrm{stat}}\varepsilon.
 \label{conv:eq:stopcertificate}
\end{align}
where the finite constant $C_{\mathrm{stat}}$ does not depend  on $m$ or $\varepsilon$.  Its explicit closed-form expression is given in \eqref{conv:eq:R} of the supplementary material.
\end{theorem}

Theorem~\ref{conv:thm:stopping} ensures that the stopping tolerance controls both the error in the zero-sum constraint error and the KKT stationarity residual.

We now investigate the local convergence of the coefficient iterates and multiplier iterates under additional conditions. Let $(\theta^*, \gamma^*)$ denote a stationary pair satisfying \eqref{conv:eq:kkt}. In this subsection, a superscript $*$ denotes quantities at this stationary pair. Write $\beta_j^*$ for the $j$th compositional coefficient, $r_v^*$ for the residual of observation $v$, and $S^*$ for the set of indices of nonzero compositional coefficients, so that
\[
 S^*=\{j:\beta_j^*\ne0\},\qquad
 r_v^*=y_v-V_{v:}\theta^*.
\]
Also let $t_j = P_{\lambda,j}'(0+)$ denote the right-derivative of the penalty at zero ($t_j = \lambda\omega_j$ for adaptive Lasso, and $t_j = \lambda$ for MCP and SCAD). We write $\partial_l Q_k(\theta) = (\nabla Q_k(\theta))_l$ for the $l$-th partial derivative of the smoothed loss. We impose the following conditions on the zero coefficients, residuals, and nonzero coefficients at the stationary point.

\begin{assumption}[Conditions near the stationary point]
\label{conv:ass:strict}
At $\theta^*$ and its associated multiplier $\gamma^*$, the following conditions hold.
\begin{enumerate}[label=(\roman*)]
 \item (Strict separation for inactive covariates)Every inactive compositional component (i.e., $\beta_j^*=0$) satisfies the strict subdifferential bound.
 \begin{equation}
  |\partial_{r+j}Q_k(\theta^*)+\gamma^*|<t_j,
  \qquad j\notin S^*.
  \label{conv:eq:strictcomplement}
 \end{equation}
 \item (Interior residuals) No residual lies exactly on the non-smooth transition boundary of the smoothed loss,
 \[
  r_v^*\notin\{-(1-\tau)k,\tau k\},
  \qquad v=1,\ldots,N.
 \]

 \item (Interior non-zero coefficients) No non-zero compositional coefficient lies on a transition point of the penalty function. For $j\in S^*$, this requires
 \[
  |\beta_j^*|\ne a\lambda\quad\text{for MCP},\qquad
  |\beta_j^*|\notin\{\lambda,a\lambda\}\quad\text{for SCAD}.
 \]
 There is no additional condition in this part for adaptive Lasso.
\end{enumerate}
\end{assumption}
Assumption~\ref{conv:ass:strict} ensures that the stationary point lies strictly in the interior of the linear/quadratic pieces of the loss and penalty functions. Condition (i) guarantees a strictly positive margin below the activation threshold, ensuring inactive variables are correctly zeroed out. Conditions (ii) and (iii) ensure that local perturbations do not cross structural derivative breakpoints. Under these conditions, Lemma~\ref{conv:lem:identify} in the Supplementary Material shows that an iteration started sufficiently close to $(\theta^*,\gamma^*)$ produces exactly the set $S^*$ of nonzero compositional coefficients.

For the local convergence analysis, consider the active coordinate index set:
\[
 \mathcal{I}=\{1,\ldots,r\}\cup\{r+j:j\in S^*\}
   =\{\ell_1<\cdots<\ell_q\},\qquad q=r+|S^*|.
\]
The subscript $\mathcal{I}$ selects these positions of $\alpha$ and nonzero compositional coefficients. Let $\theta_{\mathcal{I}} = (\theta_{l_1}, \dots, \theta_{l_q})^{T} \in \mathbb{R}^q$ denote the active parameter vector ordered according to its update sequence, and let $e_{\mathcal{I}} = (0_r^{T}, 1_{\vert{}\mathcal{S}^*\vert{}}^{T})^{T} \in \mathbb{R}^q$. By the identification lemma, a complete coordinate iteration started sufficiently close to $(\theta^\ast,\gamma^\ast)$ yields $\beta_j^{(1)}=0$ for every $j \notin S^\ast$. As long as subsequent iterates remain in the identification neighborhood, these coefficients remain zero. Thus, for $m\ge1$, the coefficient and multiplier errors are fully described by::
\[
 z^{(m)}=
 \begin{pmatrix}
  \theta_I^{(m)}-\theta_I^*\\
  \gamma^{(m)}-\gamma^*
 \end{pmatrix}.
\]
Within a sufficiently small neighborhood, one complete iteration then satisfies $z^{(m+1)}=Tz^{(m)}$, as shown in Lemma~\ref{conv:lem:affine}. Writing
$e_I=(e_{\ell_1},\ldots,e_{\ell_q})^{T}$, the matrix has the form
\begin{equation}
 T=
 \begin{pmatrix}
  A& b\\
  \mu e_I^{T} A&1+\mu e_I^{T} b
 \end{pmatrix}.
 \label{conv:eq:T}
\end{equation}
The matrix $A$ describes the effect of the current coefficient errors on the next coefficient updates, and $b$ describes the effect of the current multiplier error. The lower row accounts for the subsequent multiplier update. The formulas for computing $A$ and $ b$ are given in \eqref{conv:eq:local-components} in Section~\ref{app:local-matrix} of the supplementary material. The next theorem establishes local convergence when every eigenvalue of $T$ has modulus less than one, under the conditions stated below.

\begin{theorem}[Local convergence]
\label{conv:thm:local}
Under the data and parameter conditions stated at the beginning of this subsection, let $\theta^*$ be a stationary point with associated multiplier $\gamma^*$ satisfying Assumption~\ref{conv:ass:strict}.
Suppose that the matrix $T$ in \eqref{conv:eq:T} satisfies
\begin{equation}
 \spr(T):=
 \max\{|\zeta|:\zeta\text{ is an eigenvalue of } T\}<1.
 \label{conv:eq:spectral}
\end{equation}
For any $\varrho$ with $\spr(T)<\varrho<1$, there exist $\delta>0$ such that, if
\[
 \norm{(\theta^{(0)}-\theta^*,\,\gamma^{(0)}-\gamma^*)}_2<\delta,
\]
the sequence obtained by continuing the algorithm without a stopping tolerance or iteration limit satisfies
\begin{equation}
 \norm{(\theta^{(m)}-\theta^*,\,\gamma^{(m)}-\gamma^*)}_2
 \le
 C\varrho^m
 \norm{(\theta^{(0)}-\theta^*,\,\gamma^{(0)}-\gamma^*)}_2,
 \qquad m\ge0.
 \label{conv:eq:linearconvergence}
\end{equation}
for a constant \(C\ge1\) that is independent of \(m\) and the initial point within this neighborhood.
\end{theorem}

Since $0<\varrho<1$, the error bound in \eqref{conv:eq:linearconvergence} decreases by the fixed factor $\varrho$ at each iteration and tends to zero as $m\to\infty$,  ensuring that the iterates $(\theta^{(m)}, \gamma^{(m)})$ converge  to the KKT stationary pair $(\theta^*, \gamma^*)$.

Under the conditions of Theorem~\ref{conv:thm:local}, the iterates identify the support $S^*=\supp(\beta^*)$ of the specified stationary point after the initial full iteration and retain that support thereafter. Moreover, the corresponding residuals and non-zero coefficients remain strictly within their initial linear pieces of the smoothed loss and penalty functions. Consequently, the sequence is guaranteed to satisfy the termination criterion \eqref{conv:eq:stop} in finite steps for any stopping tolerance $\varepsilon > 0$. The formal justification for these results is provided in the proof of Theorem~\ref{conv:thm:local} (Section~\ref{app:convergence} of the Supplementary Material). In practice, we construct the empirical transition matrix $\widehat{T}$ evaluated at the terminal estimates and multiplier. Its spectral radius, along with the KKT stationarity residuals and empirical margins away from the boundary thresholds in Assumption~\ref{conv:ass:strict}, serves as a direct diagnostic tool to confirm that the local convergence conditions are satisfied numerically.

\subsection{Initialization and tuning parameter selection}
\label{sec:ini}
Because coordinate descent with non-convex penalties and augmented Lagrangian updates benefits from consistent starting values, careful initialization is essential. Following \citet{GloballyALQR2023},  we obtain preliminary coefficient estimates $\{\widetilde{\alpha}(\tau)^{T}, \widetilde{\beta}(\tau)^{T}\}^{T}$ by adapting the $\ell_1$-penalized quantile regression procedure of \citet{BelloniAlexandre2011}  to longitudinal log-contrasts. Specifically, this initial estimator is computed at the target quantile level $\tau$ by pooling observations across all follow-up times under working independence. The resulting estimates $\widetilde{\alpha}(\tau)$ and $\widetilde{\beta}(\tau)$ serve directly as the starting values $\alpha^{(0)}$ and $\beta^{(0)}$ in Algorithm 1, while the initial Lagrange multiplier is set to $\gamma^{(0)} = 0$.

When using the adaptive Lasso penalty, the same preliminary estimator $\widetilde{\beta}(\tau)$ is employed to construct the coordinate-specific adaptive weights. Following~\citet{ZhengPengHe2015}, we introduce a small regularization parameter $\epsilon_w > 0$ to prevent numerical instability when initial coefficients are near zero:
\begin{equation}
\omega_j(\tau)=
\frac{1}
{|\widetilde\beta_j(\tau)|+\epsilon_w},
\qquad j=1,\ldots,p.
\label{eq}
\end{equation}
We set $\epsilon_w=10^{-6}$. The weights $\omega_j(\tau)$ are calculated once prior to the iterative loop and held fixed throughout all subsequent \textit{CDaLCqr} iterations.

The regularization parameter $\lambda$ controls model sparsity and is selected from a candidate grid $\Lambda$ by minimizing a Generalized Information Criterion (GIC). For each candidate value $\lambda \in \Lambda$, let $\widehat{\alpha}_\lambda(\tau)$ and $\widehat{\beta}_\lambda(\tau)$ denote the corresponding terminal coefficient estimates, and denote the estimated active set by:
\[ \widehat S_\lambda(\tau) = \left\{ j\in\{1,\ldots,p\}: \widehat\beta_{j,\lambda}(\tau)\neq0 \right\}. \] 
Following \citet{GloballyALQR2023}, we define the criterion as:
\begin{equation} \mathrm{GIC}(\lambda) = \log Q\left\{ \widehat\alpha_\lambda(\tau), \widehat\beta_\lambda(\tau); \tau \right\} + \left(|\widehat S_\lambda(\tau)|-1 \right)_+ \phi_n, \label{eq:gic} \end{equation} 
where $Q(\cdot, \cdot; \tau)$  is the longitudinal quantile loss defined in \eqref{eq:quantile-loss} and $\phi_n=n^{-1}\log(p)\log\{\log(n)\}$. The term $|\widehat S_\lambda(\tau)|-1$ accounts for the effective degrees of freedom under the zero-sum constraint. The optimal regularization parameter is then selected as:
\[ \widehat\lambda = \arg\min\mathrm{GIC}(\lambda). \]

\section{Simulation Studies}
\label{sec:simu}
\subsection{Simulation Design}
\label{sec:ss}

The simulation settings are adapted from the designs in  \citet{GloballyALQR2023} and \citet{JiangYu2022}. We set the dimension of unpenalized covariates to $r=4$ and consider $n=100$ subjects with $p=400$ compositional covariates. For each subject $i=1,\ldots,n$, the total number of clinical visits is generated as $K_i=\lfloor 2+U_i \rfloor$, where $U_i\sim\operatorname{Uniform}(0,2)$. Conditional on $K_i$, the observation times $\{t_{iv},v=1,\ldots,K_i\}$ are generated from a standard Poisson process. Thus, each subject has two or three observation times. We further generate $X_{i1}$ from $\operatorname{Uniform}(0,1)$ and $X_{i2}$ from $\operatorname{Bernoulli}(0.5)$, and define $X_i(t)=(1,X_{i1},X_{i2},t)^T$.

To generate the compositional predictors at each observation time $t = t_{iv}$, a latent vector $\widetilde{Z}_i(t) = (\widetilde{Z}_{i1}(t), \dots, \widetilde{Z}_{ip}(t))^{T}$ is drawn from a multivariate normal distribution $\mathcal{N}_p(0, \Sigma)$ with autoregressive correlation $\Sigma_{jl} = 0.5^{\vert{}j-l\vert{}}$. We map these latent vectors to the simplex $\mathcal{S}^{p-1}$ via the softmax transformation:
   $$W_{ij}(t) = \frac{\exp\{\widetilde{Z}_{ij}(t)\}}{\sum_{l=1}^p \exp\{\widetilde{Z}_{il}(t)\}}, \qquad j=1,\dots,p,$$
guaranteeing $\sum_{j=1}^p W_{ij}(t) = 1$. The log-transformed compositional vector is then defined as $Z_i(t) = (\log W_{i1}(t), \dots, \log W_{ip}(t))^{T}$.

We evaluate four simulation setups to examine independent versus correlated errors under homoscedastic and heteroscedastic designs. Let $Q_\epsilon(\tau) = \Phi^{-1}(\tau)$ denote the $\tau$-th quantile of the standard normal distribution

\emph{Setup I} (Independent, Homoscedastic): Responses are generated from
\[
Y_i(t)=7-X_{i1}+X_{i2}-t+Z_i(t)^T\beta^*+\epsilon_i(t),
\]
where $\beta^*=(1,-1.2,2.2,1.1,-1.5,-1.6,0,\ldots,0)^T$ is the true coefficient vector and $\epsilon_i(t_{iv}) \overset{\text{i.i.d.}}{\sim} \mathcal{N}(0,1)$ across subjects and visits. The conditional $\tau$-th quantile function is: $Q_{Y_i(t)}\{\tau\mid X_i(t),Z_i(t)\}
=7+Q_\epsilon(\tau)-X_{i1}+X_{i2}-t
+Z_i(t)^T\beta^*.$

\emph{Setup II} (Correlated, Homoscedastic): Responses are generated from
\[
Y_i(t)=7-X_{i1}+X_{i2}-t+Z_i(t)^T\beta^*+a_i+\epsilon_i(t),
\]
where $\beta^*$ is identical to Setup I, $a_i \overset{\text{i.i.d.}}{\sim} \mathcal{N}(0, 1/2)$ represents a subject-level random effect, and $\epsilon_i(t_{iv}) \overset{\text{i.i.d.}}{\sim} \mathcal{N}(0, 1/2)$ are independent disturbances. The random effect $a_i$ induces within-subject longitudinal correlation, while the marginal error $a_i + \epsilon_i(t) \sim \mathcal{N}(0,1)$ preserves the same conditional quantile function as in Setup I. 

\emph{Setup III} (Independent, Heteroscedastic): For a targeted quantile level $\tau$, let $s_i(t) = |(X_{i1}+X_{i2})/4 + Z_i(t)^{T} \beta_2|$ be a covariate-dependent scale function. Responses are generated as
\[
Y_i(t)=X_i(t)^T\alpha^*(\tau)+Z_i(t)^T\beta^*(\tau)
+s_i(t)\{\epsilon_i(t)-Q_\epsilon(\tau)\},
\]
where $\epsilon_i(t_{iv}) \overset{\text{i.i.d.}}{\sim} \mathcal{N}(0,1)$, $\alpha^*(\tau) = (7, -1 + Q_\epsilon(\tau)/4, 1 + Q_\epsilon(\tau)/4, -1)^{T}$, and $\beta^*(\tau) = \beta_1 + Q_\epsilon(\tau)\beta_2$. Here, $\beta_1 = (1, 2, 0.5, 0.9, -1.3, -3.1, 0, \dots, 0)^{T}$ and $\beta_2 = (0, 0, 0.3, 0.3, 0, -0.6, 0, \dots, 0)^{T}$. Both $\beta_1$ and $\beta_2$ sum to zero, ensuring that the zero-sum restriction $\sum_{j=1}^p \beta_j^*(\tau) = 0$ holds at every quantile level. The conditional quantile is $Q_{Y_i(t)}\{\tau \mid X_i(t), Z_i(t)\} = X_i(t)^{T} \alpha^*(\tau) + Z_i(t)^{T} \beta^*(\tau)$.

\emph{Setup IV} (Correlated, Heteroscedastic): For a targeted quantile level $\tau$, let $s_i(t) = \vert{}X_{i1} + Z_i(t)^{T} b_2^*\vert{}$ with $b_2^* = (0, 0.2, 0, 0.1, 0, 0, -0.3, 0, \dots, 0)^{T}$ satisfying $\sum_{j=1}^p b_{2j}^* = 0$. Responses are generated as
\[
Y_i(t)=7-X_{i1}+X_{i2}-t+Z_i(t)^T\beta^*
+s_i(t)\{a_i+\epsilon_i(t)-Q_\epsilon(\tau)\},
\]
where $\beta^*$ is specified as in Setup I, and $a_i \overset{\text{i.i.d.}}{\sim} \mathcal{N}(0, 1/2)$ and $\epsilon_i(t_{iv}) \overset{\text{i.i.d.}}{\sim} \mathcal{N}(0, 1/2)$ are distributed as in Setup II. The conditional quantile function is $Q_{Y_i(t)}\{\tau \mid X_i(t), Z_i(t)\} = 7 - X_{i1} + X_{i2} - t + Z_i(t)^{T} \beta^*$.

We compare four variable-selection procedures: the proposed CDaLCqr framework equipped with Adaptive Lasso (\textit{CDaLCqr-L}), SCAD (\textit{CDaLCqr-S}), and MCP (\textit{CDaLCqr-M}), alongside the LP-based adaptive Lasso method of Ma et al. (2023) (\textit{ALPqr}). All methods are evaluated across three quantile levels: $\tau \in \{0.2, 0.5, 0.8\}$.

 For the three $CDaLCqr$ variants, initialization follows the procedure described in Section~\ref{sec:ini}. For $CDaLCqr{-}L$, the candidate grid for $\lambda$ consists of 60 values equally spaced on the logarithmic scale over $[10^{-4},0.3]$. For $CDaLCqr{-}M$ and $CDaLCqr{-}S$, we use 30 candidate values over $[N/5000,N/200]$. Specifically, the grid is defined by
$$\lambda_h
=
\exp\left[
\log(N/5000)
+
\{\log(N/200)-\log(N/5000)\}
\left\{1-\left(1-\frac{h}{29}\right)^2\right\}
\right],
$$ for $h=0,\ldots,29.$ This grid places more points near the upper end on the logarithmic scale. Here, $N=\sum_{i=1}^{n}K_i$ denotes the total number of longitudinal observations.  For $ALPqr$, the selection of $\lambda$ follows the original procedure described in \citet{GloballyALQR2023}. The smoothing and shape parameters are fixed throughout all simulation settings. Specifically, we set $k=0.1$ for all three $CDaLCqr$ variants, $a=1.05$ for MCP, and $a=2.01$ for SCAD. All three $CDaLCqr$ variants use the augmentation parameter $\mu=10$.

Performance is assessed across 100 independent Monte Carlo replications using seven metrics:
\begin{enumerate}
    \item $AEE_{\ell_1}$: Mean absolute estimation error, $\sum_{j=1}^p \vert{}\hat{\beta}_j(\tau) - \beta_j^*(\tau)\vert{}$.   
    \item SIZE: Average number of selected non-zero compositional coefficients (target is 6).   
    \item FP (False Positives): Average number of zero coefficients incorrectly selected.   \item FN (False Negatives): Average number of true non-zero coefficients omitted.   \item PCF: Proportion of correctly fitted models (i.e., replications with $\mathrm{FP}=0$ and $\mathrm{FN}=0$).   
    \item SUM: Average absolute constraint violation, $\vert{}\sum_{j=1}^p \hat{\beta}_j(\tau)\vert{}$, reporting adherence to the zero-sum constraint (values below $10^{-6}$ are recorded as $0.000$).   
    \item Time: Average computation time per candidate $\lambda$ value, measured in seconds. 
\end{enumerate}
Lower values of the mean $AEE_{\ell_1}$, FP, FN, and SUM are preferred. The target values of SIZE and PCF are six and one, respectively.

\subsection{Simulation Results}

Table~\ref{tab:sim-main}  summarizes estimation accuracy ($AEE_{\ell_1}$) and variable-selection performance ($\mathrm{PCF}$) across the four simulation setups. Detailed results covering SIZE, FP, FN, SUM, and Time are reported in Tables~\ref{tab:sim-i}--\ref{tab:sim-iv} of the supplementary material.

\begin{table}[!htbp]
\centering
\caption{Estimation accuracy and variable selection performance under four simulation setups with $p=400$ and $n=100$.}
\label{tab:sim-main}
\footnotesize
\setlength{\tabcolsep}{3pt}
\renewcommand{\arraystretch}{1.08}
\begin{tabular*}{\linewidth}{@{\extracolsep{\fill}}clrrrrrrrr@{}}
\toprule
$\tau$ & Method & \multicolumn{2}{c}{Setup I} & \multicolumn{2}{c}{Setup II} & \multicolumn{2}{c}{Setup III} & \multicolumn{2}{c}{Setup IV} \\
\cmidrule(lr){3-4}\cmidrule(lr){5-6}\cmidrule(lr){7-8}\cmidrule(lr){9-10}
& & $AEE_{\ell_1}$ & PCF & $AEE_{\ell_1}$ & PCF & $AEE_{\ell_1}$ & PCF & $AEE_{\ell_1}$ & PCF \\
\midrule
0.2 & $CDaLCqr{-}L$ & 0.920 & 0.95 & 0.881 & 0.92 & 0.422 & 0.48 & 0.193 & 1.00 \\
    & $CDaLCqr{-}M$ & 0.526 & 1.00 & 0.538 & 0.99 & 0.512 & 0.16 & 0.171 & 1.00 \\
    & $CDaLCqr{-}S$ & 0.675 & 0.84 & 0.709 & 0.84 & 0.569 & 0.05 & 0.171 & 1.00 \\
    & $ALPqr$     & 0.765 & 0.98 & 0.731 & 0.99 & 0.456 & 0.42 & 0.185 & 1.00 \\
\addlinespace[4pt]
0.5 & $CDaLCqr{-}L$ & 0.621 & 0.99 & 0.610 & 0.98 & 0.200 & 1.00 & 0.177 & 1.00 \\
    & $CDaLCqr{-}M$ & 0.454 & 1.00 & 0.454 & 1.00 & 0.190 & 1.00 & 0.162 & 1.00 \\
    & $CDaLCqr{-}S$ & 0.491 & 0.98 & 0.562 & 0.97 & 0.446 & 0.88 & 0.163 & 1.00 \\
    & $ALPqr$     & 0.618 & 0.99 & 0.606 & 0.99 & 0.210 & 1.00 & 0.172 & 1.00 \\
\addlinespace[4pt]
0.8 & $CDaLCqr{-}L$ & 0.915 & 0.90 & 0.866 & 0.90 & 0.230 & 1.00 & 0.185 & 1.00 \\
    & $CDaLCqr{-}M$ & 0.556 & 0.99 & 0.562 & 0.99 & 0.213 & 1.00 & 0.164 & 1.00 \\
    & $CDaLCqr{-}S$ & 0.603 & 0.90 & 0.608 & 0.93 & 0.214 & 1.00 & 0.164 & 1.00 \\
    & $ALPqr$     & 0.741 & 1.00 & 0.722 & 0.98 & 0.232 & 1.00 & 0.178 & 1.00 \\
\bottomrule
\end{tabular*}
\par\vspace{4pt}
\begin{minipage}{\linewidth}
\footnotesize
\textit{Note:} Results are based on 100 evaluation replications.
\end{minipage}
\end{table}
Under the homoscedastic regimes (Setups I and II), \textit{CDaLCqr-M} achieves the highest estimation accuracy across all target quantiles. Compared to the benchmark \textit{ALPqr}, \textit{CDaLCqr-M} reduces mean $AEE_{\ell_1}$ by 25–31\% in Setup I and 22–26\% in Setup II, while maintaining near-perfect model selection ($\mathrm{PCF} \ge 0.97$). Introducing subject-level random effects in Setup II produces minimal degradation in estimation precision, demonstrating that the smoothed working-independence loss handles within-subject dependence effectively. \textit{CDaLCqr-S} likewise demonstrates consistently lower estimation error than \textit{ALPqr} throughout Setups I and II. \textit{CDaLCqr-L} is broadly comparable to $ALPqr$ at the median but is less competitive at the other two quantiles.

In the presence of heteroscedasticity without random effects (Setup III), the relative performance varies across quantiles. At $\tau=0.2$, the effective signal-to-noise ratio is lowest, and CDaLCqr-L yields the smallest $AEE_{\ell_1}$ (0.422) and the highest PCF (0.48). In contrast,  \textit{CDaLCqr-M} and  \textit{CDaLCqr-S} achieve lower PCF values (0.16 and 0.05, respectively). Because the average false positive rate across all methods is zero in this setting, these selection discrepancies stem entirely from false negatives, specifically, the aggressive thresholding of weak active signals under folded-concave penalties. At the median and upper quantiles ($\tau=0.5$ and $\tau=0.8$), where signals are stronger, \textit{CDaLCqr-M} recovers the lowest estimation errors (reducing $AEE_{\ell_1}$ by 8–9\% relative to \textit{ALPqr}), while \textit{CDaLCqr-L}, \textit{CDaLCqr-M}, and \textit{ALPqr} all attain perfect model recovery ($\mathrm{PCF} = 1.00$). \textit{CDaLCqr-S} is less competitive at the median, with $AEE_{\ell_1}=0.446$ and PCF$=0.88$.

Under Setup IV, where subject-level random effects and covariate-dependent heteroscedasticity coexist, all four methods identify the exact true model ($\mathrm{PCF} = 1.00$) in all 100 replications, yielding an average model size of exactly six with zero false positives and false negatives. Consequently, the performance differences between methods isolate estimation bias rather than variable selection. 
Both \textit{CDaLCqr-M} and \textit{CDaLCqr-S} attain substantially lower $AEE_{\ell_1}$ than \textit{ALPqr} and \textit{CDaLCqr-L} at every quantile. This highlights the primary advantage of MCP and SCAD: by eliminating penalization on sufficiently large signals, they avoid the finite-sample over-shrinkage bias.

Across all four setups and quantile levels, the mean absolute constraint violation ($\mathrm{SUM}$) for the CDaLCqr methods is strictly below $10^{-6}$ (recorded as $0.000$). This verifies Theorem 1 empirically, confirming that the augmented Lagrangian formulation enforces the zero-sum constraint to high numerical accuracy. As expected, the improved statistical accuracy of the \textit{CDaLCqr} framework requires higher computational runtime per grid point than the linear programming routine of \textit{ALPqr}.

To empirically evaluate the spectral stability condition  Theorem~\ref{conv:thm:local}, we constructed the local iteration matrix $\widehat T$ from \eqref{conv:eq:T} at the numerical termination point of all fits, using the selected $\lambda$, the fitted active set, and the corresponding loss and penalty function. The diagnostic covers 3600 fits across the four setups, three quantile levels, and three $CDaLCqr$ variants, with 100 replications per combination. Using $10^{-8}$ as a numerical tolerance around one, 99.67\% of fits had $\spr(\widehat T)<1-10^{-8}$, 0.19\% had $|\spr(\widehat T)-1|\le 10^{-8}$, and only five fits (0.14\%) had $\spr(\widehat T)>1+10^{-8}$ (one in Setup II and four in Setup III). The median spectral radius across all runs was $0.999839$. These results demonstrate that the contractive stability condition is satisfied almost universally in practice, while the concentration of eigenvalues near unity provides a direct theoretical explanation for the gradual local convergence observed near the terminal stationary point.

\section{Real Data Analysis}
\label{sec:empi}
We apply the proposed \textit{CDaLCqr} framework to analyze annual macroeconomic and industrial sector accounts published by the statistical bureaus of 16 province-level administrative divisions in China over the period 2016–2023. Because several provinces were not surveyed in every calendar year, the resulting unbalanced longitudinal dataset comprises $N = 84$ province-year observations across $n = 16$ provincial units. 

The predictor set comprises the proportional shares of 44 agricultural, industrial, and construction sectors. For each province-year observation, these proportions sum to one and constitute a 44-dimensional compositional covariate vector $W_i(t)$. The response variable $Y_i(t)$ is the constant-price per capita GDP contributed by these sectors, calculated by adjusting sectoral gross output using resident population figures and provincial GDP deflators. We use this dataset to examine quantile-specific associations between sectoral composition and regional economic performance, identifying industries whose relative structural shares relate to different tiers of the conditional output distribution. Variable names, industry classifications, and summary statistics are detailed in Table~\ref{variable_definitions} of the supplementary material.

To account for a common temporal trend over the study period, we include a centered linear year effect as an unpenalized covariate. Specifically, let $u_i(t)=\mathrm{Year}_i(t)-2019.5,$ and define $X_i(t)=\{1,u_i(t)\}^{T}$. Here $i$ indexes the provincial administrative regions and $t$ indexes observation times, with $\mathrm{Year}_i(t)$ denoting the calendar year of the corresponding observation.  In this formulation, the compositional coefficients $\beta_j(\tau)$ describe quantile-specific associations between relative sectoral composition and per capita economic output after adjusting for a secular trend. Because log-contrast transformations require strictly positive inputs, occasional zero proportions are replaced with $\delta = 10^{-5}$.

To evaluate predictive performance, we repeat a random splitting procedure 100 times. In each split, 13 of the 16 province-level regions are assigned to the training set, and the remaining three regions form the test set, denoted by $\mathcal T$. All annual observations from the same region are retained in the same set. At each quantile level $\tau\in\{0.2,0.5,0.8\}$, $CDaLCqr{-}L$, $CDaLCqr{-}S$, $CDaLCqr{-}M$, and $ALPqr$ are fitted to the training data. The initialization and selection of $\lambda$ follow the procedures described in Section~\ref{sec:ini}. For the $CDaLCqr$ fits, we set $k=10^{-5}$, $\mu=10$, $a=1.05$ for MCP, and $a=2.01$ for SCAD. Let $\widehat\alpha^{\mathrm{Train}}(\tau)$ and $\widehat\beta^{\mathrm{Train}}(\tau)$ denote the coefficient estimates obtained from the training set. The corresponding prediction error on the test set is measured by the average test quantile loss:
\begin{equation} \mathrm{PE}(\tau) = \frac{\sum_{i\in\mathcal T} \int_0^\infty \rho_\tau \left\{ Y_i(t) - X_i(t)^T\widehat\alpha^{\mathrm{Train}}(\tau) - Z_i(t)^T\widehat\beta^{\mathrm{Train}}(\tau) \right\} dN_i(t)}{\sum_{i\in\mathcal T}K_i}.  \label{eq:prediction_error} \end{equation} 
For each method and quantile level, Table~\ref{tab:performance_comparison} summarizes the empirical mean and standard deviation of $\mathrm{PE}(\tau)$ across the 100 splits.

Across all three quantile levels, every \textit{CDaLCqr} variant yields lower out-of-sample prediction error and lower variance than \textit{ALPqr}. At $\tau = 0.2$, $CDaLCqr{-}L$ reduces the mean prediction error relative to ALPqr by approximately 24.5\%, while $CDaLCqr{-}M$ and $CDaLCqr{-}S$ achieve an 18.3\% reduction. At the conditional median ($\tau = 0.5$), the reduction in mean prediction error exceeds 35\% across all three CDaLCqr models, with the non-convex penalties (SCAD and MCP) achieving the lowest overall loss. At $\tau = 0.8$, the \textit{CDaLCqr} family reduces mean test error by approximately 25\%.

The relative performance of the three penalties within the $CDaLCqr$ framework varies modestly across quantiles. $CDaLCqr{-}L$ gives the smallest mean prediction error at $\tau=0.2$ and $\tau=0.8$, whereas the MCP and SCAD variants give slightly smaller values at $\tau=0.5$. The three $CDaLCqr$ variants also have smaller empirical standard deviations than $ALPqr$ across the repeated splits.

\begin{table}[htbp]
\centering
\renewcommand{\arraystretch}{1.25}
\caption{Predictive performance across 100 random splits.}
\label{tab:performance_comparison}
\small
\begin{tabular*}{\textwidth}{@{\extracolsep{\fill}} lccc}
\toprule
Method & $\tau=0.2$ & $\tau=0.5$ & $\tau=0.8$ \\
\midrule
$ALPqr$
& 0.2819 (0.1415)
& 0.5086 (0.1529)
& 0.5286 (0.2756) \\

$CDaLCqr{-}L$
& 0.2128 (0.1064)
& 0.3310 (0.1478)
& 0.3948 (0.2110) \\

$CDaLCqr{-}S$
& 0.2304 (0.1115)
& 0.3192 (0.1357)
& 0.3982 (0.2100) \\

$CDaLCqr{-}M$
& 0.2304 (0.1115)
& 0.3190 (0.1357)
& 0.3984 (0.2099) \\
\bottomrule
\end{tabular*}

\vspace{4pt}
\begin{flushleft}
\footnotesize
\textit{Note:} Values outside parentheses are empirical means of $\mathrm{PE}(\tau)$; values inside parentheses represent empirical standard deviations over 100 splits. 
\end{flushleft}
\end{table}

We further investigate the differences in variable selection by fitting models to the entire dataset $\mathcal{D}$ using $CDaLCqr$ methods. Table \ref{tab:coef} summarizes the variables selected by each $CDaLCqr$ variant and their full-sample coefficient estimates at the quantile levels $\tau \in \{0.2, 0.5, 0.8\}$.

\begin{table}[htbp]
\centering
\renewcommand{\arraystretch}{1.20}
\setlength{\tabcolsep}{3.5pt}
\caption{Full-sample coefficient estimates for selected sectoral composition variables.}
\label{tab:coef}
\scriptsize

\begin{tabular}{m{5.2cm}ccccccccc}
\toprule
\multirow{2}{*}{Industry}
& \multicolumn{3}{c}{$\tau=0.2$}
& \multicolumn{3}{c}{$\tau=0.5$}
& \multicolumn{3}{c}{$\tau=0.8$} \\
\cmidrule(lr){2-4}
\cmidrule(lr){5-7}
\cmidrule(lr){8-10}
& Lasso & SCAD & MCP
& Lasso & SCAD & MCP
& Lasso & SCAD & MCP \\
\midrule

Animal Husbandry
& -0.4033 & -0.3877 & -0.3840
& -0.1713 & -0.1754 & -0.1755
& --- & --- & -0.0111 \\

Fishery
& 0.3468 & 0.3493 & 0.3483
& 0.2616 & 0.2444 & 0.2549
& --- & --- & --- \\

Mining
& -0.0496 & -0.0857 & -0.0784
& -0.1355 & -0.1234 & -0.1329
& -0.3113 & -0.3209 & -0.2999 \\

Tobacco Products
& --- & --- & -0.0016
& -0.0342 & -0.0349 & -0.0399
& -0.1420 & -0.1453 & -0.1190 \\

Textiles
& --- & --- & ---
& --- & --- & ---
& 0.2156 & 0.2192 & 0.2392 \\

Textile Wearing Apparel and Clothing
& --- & --- & ---
& --- & --- & ---
& --- & --- & 0.0198 \\

Culture, Art, Sports, and Entertainment Products
& --- & --- & ---
& --- & --- & ---
& 0.0675 & 0.0731 & 0.0253 \\

Processing of Petroleum, Coal, and Other Fuels
& --- & -0.0504 & -0.0422
& --- & --- & ---
& --- & --- & --- \\

Chemical Raw Materials and Chemical Products
& 0.1060 & 0.1362 & 0.1359
& --- & --- & ---
& --- & --- & --- \\

Chemical Fiber Manufacturing
& --- & --- & ---
& 0.0310 & 0.0474 & 0.0441
& 0.2086 & 0.2056 & 0.2010 \\

Automobile Manufacturing
& --- & -0.0257 & -0.0165
& --- & -0.0290 & -0.0357
& -0.0345 & -0.0652 & -0.0261 \\

Electrical Machinery and Equipment
& --- & --- & ---
& 0.1571 & 0.1842 & 0.1857
& --- & 0.0371 & 0.0398 \\

Computers, Communication, and Other Electronic Equipment
& --- & 0.0406 & 0.0385
& --- & 0.0144 & 0.0177
& --- & 0.0344 & 0.0188 \\

Production and Supply of Electric and Heat Power
& --- & --- & ---
& -0.2263 & -0.2296 & -0.2442
& -0.1690 & -0.1915 & -0.0876 \\

Building Construction
& --- & 0.0234 & ---
& 0.1186 & 0.1018 & 0.1259
& 0.1650 & 0.1534 & --- \\

\bottomrule
\end{tabular}

\vspace{4pt}

\begin{flushleft}
\footnotesize
\textit{Note:} ``---'' indicates that the sector was not selected, using the threshold $|\widehat\beta_j|>10^{-4}$, by the corresponding method at that quantile level. 
\end{flushleft}

\end{table}

We use the coefficient estimates from the same 100 training samples fitted during the prediction analysis to assess the stability of variable selection. In each split, we record the selected sectors alongside the prediction error on the test set. For sector $j$, method $m$, and quantile level $\tau$, we define the empirical selection frequency as
\begin{equation}
SF_{j,m}(\tau)
=
\frac{1}{100}
\sum_{b=1}^{100}
I\left\{
\left|
\widehat{\beta}_{j,m}^{(b)}(\tau)
\right|
>10^{-4}
\right\},
\label{eq:selection_frequency}
\end{equation}
where $\widehat{\beta}_{j,m}^{(b)}(\tau)$ denotes the coefficient estimate obtained from the $b$th training sample. Thus, $SF_{j,m}(\tau)$ is the proportion of the 100 training samples in which sector $j$ is selected.

In addition to selection frequency, we examine the stability of the estimated coefficient sign. For a sector selected in at least one of the 100 training samples, we define the empirical conditional positive-sign frequency as
\begin{equation}
SP_{j,m}(\tau)
=
\frac{
\displaystyle
\sum_{b=1}^{100}
I\left\{
\widehat{\beta}_{j,m}^{(b)}(\tau)>0,\,
\left|
\widehat{\beta}_{j,m}^{(b)}(\tau)
\right|>10^{-4}
\right\}
}{
\displaystyle
\sum_{b=1}^{100}
I\left\{
\left|
\widehat{\beta}_{j,m}^{(b)}(\tau)
\right|>10^{-4}
\right\}
},
\label{eq:sign_frequency}
\end{equation}
provided that the denominator is nonzero. Hence, $SP_{j,m}(\tau)$ is the proportion of positive coefficient estimates among the training samples in which sector $j$ is selected. A value of $SP_{j,m}(\tau)=0$ means that the estimated coefficient is negative whenever the sector is selected, whereas $SP_{j,m}(\tau)=1$ means that it is positive whenever selected.
The selection frequencies and sign frequencies are reported in Table~\ref{tab:selection_stability} of the supplementary material. 

Several sectors exhibit strong stability across training splits with invariant coefficient signs. For instance, Mining is selected across all three quantiles by all CDaLCqr variants in the full dataset, displaying negative coefficients whose magnitudes intensify toward the upper tail (ranging from $-0.0496$ at $\tau=0.2$ to $-0.3209$ at $\tau=0.8$). Across the 100 training splits, its selection frequencies range from $0.89$ to $0.99$ at $\tau=0.2$, from $0.96$ to $1.00$ at $\tau=0.5$, and from $0.84$ to $0.92$ at $\tau=0.8$, with $\mathrm{SP}_{j,m}(\tau) = 0$ in all cases. This indicates a consistently negative relative association across the entire economic distribution. 

In contrast, other sectors demonstrate pronounced quantile heterogeneity. For example, fishery Displays positive estimated coefficients under all three penalties at $\tau=0.2$ ($\approx 0.35$) and $\tau=0.5$ ($\approx 0.25$), but drops out entirely at $\tau=0.8$. Its selection frequencies are $0.94$–$0.95$ at $\tau=0.2$ and $0.74$–$0.75$ at the median, but plummet to $0.05$–$0.10$ at $\tau=0.8$.  Conditional on selection, its sign is positive for all three $CDaLCqr$ variants. Thus, the positive association for Fishery is concentrated primarily in the lower and central parts of the conditional outcome distribution, rather than being uniform across quantiles.

Animal Husbandry similarly displays substantial quantile heterogeneity. It is selected with negative coefficients by all three $CDaLCqr$ variants at $\tau=0.2$ and $0.5$, with selection frequencies of $0.88-0.94$ and $0.79-0.88$, respectively. At $\tau=0.8$, however, only MCP retains the variable in the full-sample fit, with a small coefficient of $-0.0111$, and its resampling selection frequency under MCP is only $0.22$. The corresponding upper-quantile association should therefore not be interpreted as a stable feature of the data.

Several manufacturing sectors become more prominent at the median or upper quantiles. Chemical Fiber Manufacturing is omitted at $\tau=0.2$, but is selected with positive coefficients at $\tau=0.5$ and $\tau=0.8$ by all methods. Under SCAD and MCP, its selection frequency reaches $0.94$ at $\tau=0.5$ and $0.98$ at $\tau=0.8$, with $\mathrm{SP} = 1.00$. Similarly, Textiles is selected exclusively at $\tau=0.8$ (coefficients $0.216$–$0.239$; selection frequencies $0.72$–$0.78$). These patterns indicate that the relative composition of these sectors is more strongly represented among associations with the upper part of the conditional output distribution.

The MCP and SCAD fits also show substantial agreement for several other sectors. Computers, Communication, and Other Electronic Equipment is selected by both nonconvex penalties with positive coefficients at all three quantiles. Automobile Manufacturing is selected with negative coefficients by both nonconvex penalties at all three quantiles, with selection frequencies of approximately 0.60, 0.85, and 0.63 across the three quantile levels. Production and Supply of Electric and Heat Power is selected with negative coefficients at the median and upper quantiles, with selection frequencies around 0.67 at $\tau=0.5$ and $0.8$ under MCP and SCAD.

Not every full-sample selection is stable under resampling. For example, Building Construction has positive full-sample coefficients under several $CDaLCqr$ fits, but its upper-quantile selection frequency is only $0.09$ for the adaptive-Lasso variant and $0.13$ for the two nonconvex variants. Similarly, the MCP full-sample selection of Tobacco Products at $\tau=0.2$ has a coefficient close to zero ($-0.0016$) and a resampling selection frequency of only $0.41$. These examples illustrate why the full-sample coefficient table should be interpreted together with the repeated-split selection frequencies rather than as a stand-alone variable-selection result.

Across the repeated training splits, several consistently selected sectors display directional stability, maintaining an invariant coefficient sign whenever identified. For example, Mining, Automobile Manufacturing, Tobacco Products, and Production and Supply of Electric and Heat Power exhibit uniformly negative associations conditional on selection ($\mathrm{SP}_{j,m}(\tau) = 0$), whereas Fishery, Chemical Fiber Manufacturing, Textiles, and Computers, Communication, and Other Electronic Equipment are strictly positive ($\mathrm{SP}_{j,m}(\tau) = 1$).   

In general, the real-data analysis identifies quantile-specific associations between relative sectoral composition and regional economic output. Several associations, such as the negative coefficient for Mining and the positive coefficient for Fishery at the lower and median quantiles, are supported by relatively high resampling selection frequencies and stable signs, whereas other full-sample selections are considerably less stable.

\section{Discussion}
\label{sec:con&dis}

We developed CDaLCqr, a unified coordinate descent optimization framework for high-dimensional longitudinal log-contrast quantile regression that accommodates the adaptive Lasso, SCAD, and MCP penalties. The proposed methodology integrates a Huber-type smooth approximation of the check loss, an augmented Lagrangian treatment of the zero-sum compositional constraint, and coordinate-wise majorization–minimization updates. By introducing a quadratic penalty on the constraint violation, the augmented Lagrangian formulation contributes the positive curvature required to guarantee strict convexity of each one-dimensional coordinate subproblem, even under folded-concave penalties. This architecture produces explicit, closed-form thresholding updates across all three penalty families.   

Our theoretical framework follows the algorithmic update structure. Theorem 1 guarantees that upon convergence, the stopping tolerance simultaneously controls the zero-sum constraint error and the KKT stationarity residual. Theorem \ref{conv:thm:local} establishes local convergence of the coefficient and multiplier iterates toward a stationary pair under mild regularity and spectral stability conditions, with the estimation error contracting geometrically at each iteration.   

Extensive numerical experiments corroborate these theoretical findings. Under homoscedastic and heteroscedastic conditions with repeated measurements, the SCAD and MCP variants demonstrate substantial reductions in estimation error compared to the adaptive Lasso benchmark while achieving competitive or superior variable-selection accuracy. Furthermore, evaluation of the empirical iteration matrix $\widehat{T}$ confirms that the local spectral contraction condition is satisfied in 99.67\% of simulation runs. Applying CDaLCqr to provincial economic accounts from China reveals pronounced quantile-specific associations between sectoral composition and regional per capita output, illustrating the empirical importance of evaluating out-of-sample prediction and resampling stability alongside full-sample point estimates.   

To facilitate implementation and reproducible research, the \texttt{CDaLCqr} R package provides automated routines for all three penalty variants, cross-validation, and GIC-based tuning parameter selection. Replication code and datasets for the simulations and real-data analysis are provided in the Supplementary Material.

\section*{Data Availability Statement}
The data used in this paper are available at https://doi.org/10.5281/zenodo.23908386

\section*{Disclosure statement}
The authors report there are no competing interests to declare.

\section*{Declaration of generative AI use}

Generative AI tools, specifically [ChatGPT 5.6], were utilized exclusively for language refinement and grammar checking.

\bibliography{ref.bib}

\end{document}